\documentclass[12pt]{article}
\usepackage{epsfig}
\makeatletter
\@addtoreset{equation}{section}
\makeatother

\newcommand{\ret}{\nonumber\\}

\newcommand{\abs}[1]{\left|#1\right|}

\newcommand{\rbk}[1]{\left(#1\right)}
\newcommand{\sqbk}[1]{\left[#1\right]}
\newcommand{\cbk}[1]{\left\{#1\right\}}
\newcommand{\bkt}[1]{\left\langle#1\right\rangle}

\newcommand{\ep}{\varepsilon}

\renewcommand{\phi}{\varphi}

\newcommand{\ket}[1]{|#1\rangle}
\newcommand{\bra}[1]{\langle #1|}

\newcommand{\Tr}{{\rm Tr}}

\newcommand{\SSHT}{Sato, Sekimoto, Hondou, and Takagi}

\newcommand{\sumin}{\sum_{i=1}^n}
\newcommand{\tbeta}{\tilde{\beta}}
\begin{document}

\begin{flushright}
\small
technical note
\end{flushright}

\noindent
{\Large \bf Inevitable irreversibility in 
a quantum system consisting of many non-interacting ``small'' pieces}

\bigskip
\noindent
Hal Tasaki\footnote{
Department of Physics,
Gakushuin University,
Mejiro, Toshima-ku, Tokyo 171,
JAPAN

electronic address: hal.tasaki@gakushuin.ac.jp
}

\begin{abstract}
We review the recent result of \SSHT\ \cite{SSHT} 
(cond-mat/0008393) on the irreversibility 
inevitably observed in systems consisting of 
many non-interacting ``small'' pieces.
We focus on quantum models, and supply an explicit lower bound for 
the work required to complete a cyclic process.
\end{abstract}

\section{Introduction}
In statistical physics textbooks, one often encounters systems which 
consist of macroscopic numbers of identical small parts which do not 
interact with each other.
This is not too unrealistic since there are many physical systems
(such as certain spin systems, including nuclear spin systems)
which 
are well approximated by such non-interacting models in 
some ranges of 
temperature and time.  

The question that we wish to examine here is the following:
{\em Does a system which consists 
of many non-interacting pieces behave 
as a ``healthy'' thermodynamic system?}\/
It is evident from exercises in statistical physics that the answer is 
``yes'' when only equilibrium properties are concerned.
If one focuses on certain non-equilibrium aspects, however,
the situation may be different.
In fact \SSHT\ recently proved that,
in a system which consists of many non-interacting small parts, 
 a simple quasi-static process 
involving contacts with two heat baths can never be reversible
in general.
It is remarkable that such a system fails to provide 
us with reversible 
processes, which are among the building blocks of conventional 
thermodynamics\footnote{
The comparison hypothesis as in \cite{LY}
is also violated, 
if we include a contact with a heat bath (whose temperature is chosen 
carefully so that no net energy is exchanged)
at the end of each ``adiabatic process.''
}.
\SSHT\ then raise an interesting 
question whether one can develop a new 
thermodynamic framework which is capable of describing these 
unavoidable irreversible processes.

The basic idea of \SSHT\ is indeed quite simple.
Suppose that a small piece (whose identical copies form the whole 
system)
is a quantum system with \( n \) energy levels
\( \ep_{1}, \ldots, \ep_{n} \).
We first assume that the whole system is in equilibrium with a heat 
bath at inverse temperature \( \beta \).
Then the probability of finding 
the small system in the \( i \)-th state is 
\( p_{i}=e^{-\beta \ep_{i}}/z(\beta) \) where \( z(\beta) \) is 
the partition function for the small system.
We then {\em gently}\/ decouple the system from the bath 
in such a way that the 
small system is still described by
the same probability \( p_{i} \).
Then we change a parameter in the model Hamiltonian very slowly, 
modifying the energy levels to \( \ep'_{1}, \ldots, \ep'_{n} \).
The adiabatic theorem tells us that, if the parameter change is 
sufficiently slow,
the probability of finding the small system in the \( i \)-th state 
is still given by the same\footnote{
We number the states so that \( \ep_{i}\le\ep_{i+1} \)
and \( \ep'_{i}\le\ep'_{i+1} \).
} \( p_{i} \).
But this \( p_{i} \) cannot be represented in the Gibbsian form
\( p_{i}=e^{-\beta' \ep'_{i}}/z'(\beta') \) for any \( \beta' \)
unless the 
two energy levels \( \ep_{1}, \ldots, \ep_{n} \) and
 \( \ep'_{1}, \ldots, \ep'_{n} \) satisfy a special condition.
 (See (\ref{eq:hh'cond}).)
Therefore if the system is brought into a contact with a heat bath 
after the operation, something nontrivial (which turns out to be a 
jump in the entropy) must take place, no matter 
how carefully one chooses the temperature of the bath.

Of course such a mechanism applies to any quantum systems, including 
truly macroscopic ones.
But it is wellknown that this kind of deviation from a certain 
equilibrium distribution is irrelevant in thermodynamic limits.
A crucial observation of \SSHT's 
is that the smallness of each non-interacting 
piece makes this deviation thermodynamically observable.

If one regards non-interacting systems as models of more realistic 
weakly-interacting systems, then the above irreversibility of 
\SSHT\ should be observed only for operations which take 
place in a time interval which is not long enough for the system 
to redistribute its energy between the small parts.
If the operation is slow enough and such a redistribution takes 
place, 
then the probability distribution for each part at the end of 
operation  becomes essentially Gibbsian, and the 
process may be reversible.

\section{Setup and main results}
We consider a quantum system which consists of \( N \) identical 
``small'' quantum systems which do not interact with each 
other.
A single small system has an \( n \)-dimensional Hilbert space,
and  the whole system has an \( nN \)-dimensional Hilbert 
space.

\begin{figure}
\centerline{\epsfig{file=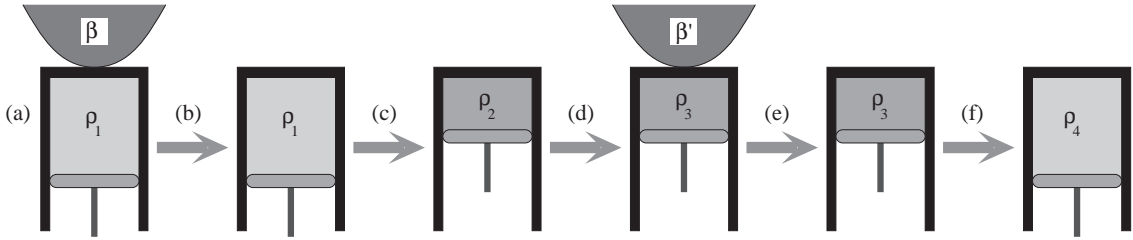,width=15cm}}
\caption[dummy]{
\SSHT\  considered the following cycle;
(a)~A macroscopic system is in touch with a heat bath at 
inverse temperature
 $\beta$.
Th state of the system is described by a density matrix  $\rho_1$.
(b)~One gently decouples the system from the heat bath,
without changing its state.
(c)~An external agent performs an operation on the system,
changing its Hamiltonian from  $H$ to  $H'$.
The state changes to  $\rho_2$, which may not be Gibbsian.
(d)~The system is put into contact with a heat bath at 
inverse temperature  $\beta'$.
Here  $\beta'$ is carefully chosen so that 
there is no net exchange of energy between the system and the bath.
Note that this is a very innocent thermal contact, which causes no observable effects in normal thermodynamic systems.
(e)~Again one gently decouples the system from the heat bath,
without changing its state.
(f)~The external agent performs an operation on the system, 
bringing back its Hamiltonian from  $H'$ to  $H$, thus completing a 
cycle.

Let  $W_{\rm cyc}$ be the total work done to the system by the external
agent during the above cycle.
Conventional thermodynamics tells us that  $W_{\rm cyc}$ 
can be made arbitrarily small by performing
the operation slowly and gently.
When the system in consideration consists of many
``small' parts which do not interact with each other, however,
\SSHT\ proved that  $W_{\rm cyc}$ is always greater than a 
finite value.
This means that the above thermodynamic cycle is 
inevitably irreversible.
}
\label{fig:cycle}
\end{figure}

The initial Hamiltonian of the system is
\begin{equation}
	H=\sum_{i=1}^N h_{i},
	\label{eq:Hinit}
\end{equation}
where, for all \( i \),
\( h_{i} \) is identical to a certain Hamiltonian \( h \)
for the small system.
Initially the system is in touch with a heat bath, and hence in 
the Gibbs state 
\begin{equation}
	\bkt{\cdots}_{1}^{\rm Gibbs}
	=\Tr[(\cdots)\rho_{1}^{\rm Gibbs}],
	\label{eq:bkt1}
\end{equation}
with
\begin{equation}
	\rho_{1}^{\rm Gibbs}
	=
	\frac{e^{-\beta H}}{\Tr[e^{-\beta H}]}.
	\label{eq:rho1i}
\end{equation}

We gradually decouple the system from the heat bath, making it 
thermally isolated.
We assume that the state of the system is unchanged from 
(\ref{eq:bkt1}).

We then perform a ``classical'' operation on the system.
This corresponds to an adiabatic\footnote{
Note that the word ``adiabatic'' means that no 
``heat'' is exchanged between the system and the outside world,
and does not mean anything like ``slow and gentle.''
} process in thermodynamics.
We model the operation by a time dependent Hamiltonian
\begin{equation}
	H(t)=\sum_{i=1}^Nh_{i}(t),
	\label{eq:H(t)}
\end{equation}
where all \( h_{i}(t) \) are identical.
After the operation, the Hamiltonian settles to 
\begin{equation}
	H'=\sum_{i=1}^Nh'_{i},
	\label{eq:H'}
\end{equation}
where all \( h'_{i} \) are identical to \( h' \).
We denote by \( U \) the unitary operator which describes the 
quantum mechanical time evolution during the whole 
operation\footnote{
\label{f:U}
The unitary operator \( U \) is formally defined as follows.
Suppose that the operation is 
done during \( t=t_{0} \) and \( t_{1} \).
Let \( U(t) \) be the solution of the Schr\"{o}dinger 
equation \( i\partial U(t)/\partial t=H(t)\,U(t) \)
with the initial condition \( U(t_{0})={\bf 1} \).
Then \( U=U(t_{1}) \).
In what follows, however, we only use the fact that 
\( U \) is unitary.
}.
Since the system is thermally isolated, 
its state after the operation is simply given by
\begin{equation}
	\bkt{\cdots}_{2}=\Tr[(\cdots)\rho_{2}],
	\label{eq:bkt2}
\end{equation}
with
\begin{equation}
	\rho_{2}=U\rho_{1}^{\rm Gibbs}U^{-1}.
	\label{eq:rho2i}
\end{equation}
Note that in general this is not a Gibbs state.

Next we put the system into a contact with another heat bath with the 
inverse temperature \( \beta' \).
After equilibration, the state of the system becomes another 
Gibbs state
\begin{equation}
	\bkt{\cdots}_{3}^{\rm Gibbs}
	=\Tr[(\cdots)\rho_{3}^{\rm Gibbs}],
	\label{eq:bkt3}
\end{equation}
with
\begin{equation}
	\rho_{3}^{\rm Gibbs}
	=
	\frac{e^{-\beta' H'}}{\Tr[e^{-\beta' H'}]}.
	\label{eq:rho3i}
\end{equation}
Here the inverse temperature is (carefully) chosen so that
\begin{equation}
	\bkt{H'}_{2}=\bkt{H'}_{3}^{\rm Gibbs}
	\label{eq:H2=H3}
\end{equation}
holds.
This means that there is no net exchange of heat between the system 
and the second heat bath\footnote{
Because the Hamiltonian (\ref{eq:H'}) is a sum of mutually commuting 
pieces, the fluctuation of the energy is \( O(\sqrt{N}) \) in the 
states  \( \bkt{\cdots}_{2} \) and
\( \bkt{\cdots}_{3}^{\rm Gibbs} \).
Since the expectation values of the energy itself is \( O(N) \),
this means that the fluctuation
of energy is simply negligible for large enough 
\( N \).
}.
This thermal contact thus looks quite innocent.
We stress that for a normal thermodynamic system,
such a thermal contact without energy exchange
produces no observable effects.
It is a peculiar nature of systems which consist of 
non-interacting small pieces that such a contact 
inevitably leads to irreversibility.

Finally we repeat the same process backwards.
We first gently decouple the system from the heat bath (without 
changing the state of the system), 
and then make an arbitrary operation which brings back the Hamiltonian 
from \( H' \) to the initial \( H \).
By denoting \( U' \) the unitary operator for this time evolution,
we can write the final (generally non-Gibbsian) state as
\begin{equation}
	\bkt{\cdots}_{4}=\Tr[(\cdots)\rho_{4}],
	\label{eq:bkt4}
\end{equation}
with
\begin{equation}
	\rho_{4}=U'\rho_{3}^{\rm Gibbs}U'^{-1}.
	\label{eq:rho4}
\end{equation}

From the energy conservation law, one sees that the difference
\( \bkt{H}_{4}-\bkt{H}_{1}^{\rm Gibbs} \)
is the total energy supplied to the system during the operation.
In sufficiently large systems where 
the energy exchange between the second heat bath and the system 
is negligible,
this difference is the total work done to the system by the 
outside agent (who 
performs the operation).
Note that, from a thermodynamic point of view,
the two operations together form a cycle because the Hamiltonian 
finally returns to the initial one.
One of the basic assumptions in conventional thermodynamics is that 
the work during such a cycle 
can be made as small as possible by performing the operation 
slowly.
The limiting operation in which the total work is vanishing is 
usually called a ``reversible'' cycle.

But the following theorem due to \SSHT\ clearly shows
that a reversible cycle is never realized in 
a system which consists of non-interacting small pieces.

\bigskip
\noindent
{\bf Theorem} (\SSHT)
For any unitary operators (which are consistent with the
given initial and 
the final Hamiltonians\footnote{
See footnote~\ref{f:U}.
}) \( U \) and \( U' \),
one has
\begin{equation}
	\bkt{H}_{4}-\bkt{H}_{1}^{\rm Gibbs}
	\ge
	N\,\Delta\ep(\beta;h,h'),
	\label{eq:HHNe}
\end{equation}
where \( \ep(\beta;h,h') \) depends only on \( \beta \) and
the energy levels \( \ep_{1},\ldots,\ep_{n} \)
and  \( \ep'_{1},\ldots,\ep'_{n} \) of the Hamiltonians
\( h \) and \( h' \), respectively.
The quantity \( \ep(\beta;h,h') \) is nonnegative in general,
and is strictly positive unless 
\begin{equation}
	\ep_{i}-\ep_{1}=A(\ep'_{i}-\ep'_{1}),
	\label{eq:hh'cond}
\end{equation}
for any \( i \) with a constant \( A \).

\bigskip
For an explicit form of \( \Delta\ep(\beta,h,h') \),
see (\ref{eq:final}).

\section{Example}
Simple (and probably realistic) 
examples are paramagnetic spin system.
Note, however, that
we do not expect the irreversibility 
due to the ``smallness'' in a paramagnetic 
\( S=1/2 \) system, where
the small system has only two levels.

The simplest example is the \( S=1 \) model 
under external magnetic field \( \bar{H} \) with
a crystal field anisotropy \( D \), whose Hamiltonian is
\begin{equation}
    H=\sum_{i=1}^N
    \cbk{
    -\bar{H} S^{(z)}_{i}
    -D (S^{(z)}_{i})^2
    }.
    \label{eq:HS=1}
\end{equation}
Here \( S^{(z)}_{i} \) is the spin operator with three eigenvalues
\( +1 \), 0, and \( -1 \).
Thus each small system has three energy levels
\( -\bar{H}-D \), 0, \( \bar{H}-D \).

As in the standard adiabatic cooling experiments,
we thermally isolate the spin system, and change the
external magnetic field \( \bar{H} \).
We find that the condition (\ref{eq:hh'cond}) is satisfied 
when \( D=0 \).
Thus we expect that the irreversibility 
due to the smallness of non-interacting pieces
is observable when \( |D/\bar{H}| \) is sufficiently
large.

\section{Proof}
\subsection{Proof of the main theorem}
We prove the main theorem by applying the ideas of
\SSHT\ to quantum models.
We stress that our argument is a straightforward implementation of that 
of \SSHT's.
The main points in the proof are 
the use of Gibbs-like entropy (\ref{eq:Sh}),  (\ref{eq:Sh'}),
and the representation (\ref{eq:S31rep}) of the entropy increase as a 
relative entropy.

Throughout the present section, we only treat a special
(and unphysical) case with \( N=1 \), i.e., 
a single small system.
Since the small systems do not interact with each other,
results for \( N=1 \) immediately implies the
corresponding results for 
general \( N \).

For \( i=1,\ldots,n \), 
let \( \ket{\phi_{i}} \) and \( \ket{\phi'_{i}} \) be
the normalized eigenstates of \( h \) and \( h' \), 
respectively.
We denote by \( \ep_{i} \) and \( \ep'_{i} \)
the corresponding energy eigenvalues.
We number the eigenstates so that
\( \ep_{i}\le\ep_{i+1} \) and 
\( \ep'_{i}\le\ep'_{i+1} \)
hold.

Let us define a unitary operator \( U \) by
\begin{equation}
	U\ket{\phi_{i}}=\ket{\phi'_{i}},
	\label{eq:Uslow}
\end{equation}
for all \( i=1,\ldots,n \).
The adiabatic theorem in quantum mechanics guarantees that the time 
evolution is described by this unitary operator when the operation 
is executed infinitely slowly (and if one tunes the phases of the 
basis states properly).
We note that, in usual macroscopic quantum systems,
an operation must be unphysically slow for the adiabatic theorem to 
be relevant.
In the present case of systems consisting of non-interacting small 
systems, however, the slowness required by the theorem may be 
physically realized since each small system evolve independently.

Once we prove the desired bounds for the special unitary 
transformation \( U \), the same bounds for arbitrary
\( U \) (and \( U' \)) follow immediately.
This is because \cite{Tasaki} for given \( h \), \( h' \), and
\( \beta \), the energy \( \bkt{h'}_{2} \) after the first operation 
for arbitrary \( U \) (consistent with the Hamiltonians 
\( h \) and \( h' \))
does not exceed the same quantity 
obtained from the unitary operation
(\ref{eq:Uslow}).
(This is nothing but the minimum work principle.)
The same is true for the second operation which brings
back the Hamiltonian from \( h' \) to \( h \).
Therefore we shall only concentrate on \( U \)
defined by (\ref{eq:Uslow}).

Let us define the Gibbs-like entropy with respect to Hamiltonian 
\( h \) as\footnote{
Note that the present definition of entropy depends on the choice of 
Hamiltonian \( h \), while the Gibbs entropy (and the von Neuman 
entropy) does not depend on Hamiltonians.
In this sense, the present definition is closer (in spirit) to the 
Boltzmann entropy, which depends on the way of characterizing the 
system from a macroscopic view point (e.g., by using energy).
}
\begin{equation}
	S_{h}[\rho]=-k\sum_{i=1}^n
	\bra{\phi_{i}}\rho\ket{\phi_{i}}
	\,
	\log \bra{\phi_{i}}\rho\ket{\phi_{i}},
	\label{eq:Sh}
\end{equation}
and the entropy with respect to \( h' \) as
\begin{equation}
	S_{h'}[\rho]=-k\sum_{i=1}^n
	\bra{\phi'_{i}}\rho\ket{\phi'_{i}}
	\,
	\log \bra{\phi'_{i}}\rho\ket{\phi'_{i}}.
	\label{eq:Sh'}
\end{equation}

Since the initial state \( \bkt{\cdots}_{1}^{\rm Gibbs} \)
is Gibbsian, we write  down the density matrix explicitly as
\begin{equation}
	\rho_{1}^{\rm Gibbs}
	=
	\frac{e^{-\beta h}}{\Tr[e^{-\beta h}]}
	=
	\sumin\ket{\phi_{i}}\,p_{i}\,\bra{\phi_{i}},
	\label{eq:rho1}
\end{equation}
where
\begin{equation}
	p_{i}=\frac{e^{-\beta\ep_{i}}}{z(\beta)},
	\quad
	z(\beta)=\sumin e^{-\beta\ep_{i}}.
	\label{eq:pi}
\end{equation}
Thus, from the definition (\ref{eq:Sh})
of entropy, we find
\begin{equation}
	S_{h}[\rho_{1}^{\rm Gibbs}]
	=
	-k\sumin p_{i}\log p_{i}
	=
	k\beta\bkt{h}_{1}+k\,\log z(\beta).
	\label{eq:Sh1}
\end{equation}

By using (\ref{eq:rho1}) and (\ref{eq:Uslow}),
we find that the density matrix after the first operation is simply 
written as
\begin{equation}
	\rho_{2}=U\rho_{1}^{\rm Gibbs}U^{-1}
	=\sumin\ket{\phi'_{i}}\,p_{i}\,\bra{\phi'_{i}},
	\label{eq:rho2}
\end{equation}
with the same \( p_{i} \) as in (\ref{eq:pi}).
We therefore find from (\ref{eq:Sh'}) that
\begin{equation}
	S_{h'}[\rho_{2}]=
	-k\sumin p_{i}\log p_{i}=
	S_{h}[\rho_{1}^{\rm Gibbs}].
	\label{eq:Sh2}
\end{equation}
Our entropy is invariant under slow operation.

We then examine the next Gibbs state 
 \( \bkt{\cdots}_{3}^{\rm Gibbs} \)
obtained by letting the system interact with the second heat bath.
Its density matrix is simply
\begin{equation}
	\rho_{3}^{\rm Gibbs}
	=
	\frac{e^{-\beta' h'}}{\Tr[e^{-\beta' h'}]}
	=
	\sumin\ket{\phi'_{i}}\,p'_{i}\,\bra{\phi'_{i}},
	\label{eq:rho3}
\end{equation}
where
\begin{equation}
	p'_{i}=\frac{e^{-\beta'\ep'_{i}}}{z'(\beta')},
	\quad
	z'(\beta')=\sumin e^{-\beta'\ep'_{i}}.
	\label{eq:pi'}
\end{equation}
Recall that the inverse temperature \( \beta' \) is
determined from the condition
\begin{equation}
	\bkt{h'}_{2}=\bkt{h'}_{3}^{\rm Gibbs}.
	\label{eq:h2=h3}
\end{equation}

From the definition (\ref{eq:Sh'}) of the entropy, 
we have
\begin{eqnarray}
	S_{h'}[\rho_{3}^{\rm Gibbs}]
	&=&
	-k\sumin p'_{i}\log p'_{i}
	\ret
	&=&
	k\beta \bkt{h'}_{3}^{\rm Gibbs}+k\,\log z'(\beta')
	\ret
	&=&
	k\beta \bkt{h'}_{2}+k\,\log z'(\beta')
	\ret
	&=&
	k\beta \sumin p_{i}\{\ep'_{i}+k\,\log z'(\beta')\}
	\ret
	&=&
	-k\sumin p_{i}\log p'_{i},
	\label{eq:Sh3}
\end{eqnarray}
where we used (\ref{eq:h2=h3}) and (\ref{eq:rho2}).
Combining this with (\ref{eq:Sh1}), we
find that the difference of the two entropies can be represented in a 
remarkable manner as
\begin{equation}
	S_{h'}[\rho_{3}^{\rm Gibbs}]-S_{h}[\rho_{1}^{\rm Gibbs}]
	=
	-k\sumin p_{i}\log\frac{p'_{i}}{p_{i}},
	\label{eq:S31rep}
\end{equation}
where the right-hand side is nothing but the relative entropy.
From the wellknown property of relative entropies,
we find\footnote{
{\em Proof:}\/
Since \( \log x\le x-1 \), one finds
\( (\mbox{L.H.S.})\ge-k\sumin p_{i}\{(p'_{i}/p_{i})-1\}
=k\sumin\{p_{i}-p'_{i}\}=0 \).
To show that the equality holds only when \( p_{i}=p'_{i} \)
for all \( i \),
it suffices to note that  \( \log x< x-1 \) whenever \( x\ne1 \).
}
\begin{equation}
	S_{h'}[\rho_{3}^{\rm Gibbs}]-
	S_{h}[\rho_{1}^{\rm Gibbs}]\ge0.
	\label{eq:S31}
\end{equation}
Here the equality holds only when  \( p_{i}=p'_{i} \)
for all \( i \).
By examining the explicit formulas (\ref{eq:pi}) and 
(\ref{eq:pi'}) of \( p_{i} \) and \( p'_{i} \), respectively,
one finds that this is possible only when the energy levels 
satisfy the condition (\ref{eq:hh'cond}).

As for the second operation (which brings back
the Hamiltonian from \( h' 
\) to \( h \)), we assume that the time evolution is described by the 
unitary operator \( U^{-1} \).
We repeat the same argument to show
\begin{equation}
	S_{h}[\rho_{4}]=S_{h'}[\rho_{3}^{\rm Gibbs}].
	\label{eq:S4S3}
\end{equation}
As a theoretical reference, we define yet another 
Gibbs state \( \bkt{\cdots}_{5}^{\rm Gibbs} \) 
with the density matrix
\begin{equation}
	\rho_{5}^{\rm Gibbs}=\frac{e^{-\beta''h}}{\Tr[e^{-\beta''h}]},
	\label{eq:rho5}
\end{equation}
where \( \beta'' \) is determined by the condition
\begin{equation}
	\bkt{h}_{5}^{\rm Gibbs}=\bkt{h}_{4}.
	\label{eq:h5h4}
\end{equation}
Then exactly the same argument as before shows
\begin{equation}
	S_{h}[\rho_{5}^{\rm Gibbs}]-
	S_{h'}[\rho_{3}^{\rm Gibbs}]\ge0.
	\label{eq:S53}
\end{equation}

From (\ref{eq:S31}) and (\ref{eq:S53}), we finally get
\begin{equation}
	S_{h}[\rho_{5}^{\rm Gibbs}]-
	S_{h}[\rho_{1}^{\rm Gibbs}]\ge0,
	\label{eq:S51}
\end{equation}
where the equality holds only when the condition 
(\ref{eq:hh'cond}) is satisfied.
Since \( \bkt{\cdots}_{1}^{\rm Gibbs} \) and
\( \bkt{\cdots}_{5}^{\rm Gibbs} \) are the Gibbs states for the same 
Hamiltonian \( h \) with the inverse temperatures \( \beta \)
and \( \beta'' \), respectively, (\ref{eq:S51}) implies
that
\begin{equation}
	\beta''\le\beta,
	\label{eq:bb}
\end{equation}
and hence
\begin{equation}
	\bkt{h}_{4}=\bkt{h}_{5}^{\rm Gibbs}
	\ge\bkt{h}_{1}^{\rm Gibbs}.
	\label{eq:h451}
\end{equation}
Again the equality holds only when 
the condition 
(\ref{eq:hh'cond}) is satisfied\footnote{
It can be shown easily that the equality indeed holds
when we have the condition 
(\ref{eq:hh'cond})
and the time evolution is described by
the special \( U \) and \( U^{-1} \)
defined by (\ref{eq:Uslow}).
For a general time evolution, the equality may be violated
for any Hamiltonians, but this is of course the 
usual irreversibility.
}.
This proves the desired theorem since, for the present choice of 
unitary operator \( U \),
the quantity \( \bkt{h}_{4}-\bkt{h}_{1}^{\rm Gibbs} \) 
clearly depends only on \( \beta \), \( \ep_1,\ldots,\ep_n \),
and  \( \ep'_1,\ldots,\ep'_n \),
and has been shown to be strictly positive unless 
(\ref{eq:hh'cond}) holds\footnote{
We note that the above proof can be directly applied to
classical systems.
To do this, we fix a constant \( \delta>0 \), and 
(as in \cite{Tasaki})
decompose the whole phase space into slices 
with the same volume \( \delta \).
The slices are determined so that the points in 
the \( i \)-th slice have initial energies 
between \( \ep_{i} \) and \( \ep_{i+1} \).
We prepare similar decomposition for the final energy as well.
Then we let \( p_{i} \) be the probability that
the initial state is found in the \( i \)-th slice.
If we perform the operation slowly,
one finds that, in the final state, 
the probability that the state is in the \( i \)-slice
(in the decomposition according to the final energy)
is still given by \( p_{i} \).
(We only use the adiabatic theorem.)
Then we can simply repeat the above proof.
By taking the limit \( \delta\to0 \) finally,
the desired proof for classical systems follow.}.

\subsection{Explicit lower bound of the work}
We stress that the argument of \SSHT\ presented in the previous 
subsection indeed proves the strict inequality
\begin{equation}
	\bkt{h}_{4}-\bkt{h}_{1}^{\rm Gibbs}>0,
	\label{eq:h41}
\end{equation}
for any \( h \) and \( h' \) which do not satisfy the
condition (\ref{eq:hh'cond}).
But it might be desirable to have an explicit positive
lower bound for the required work
\( \bkt{h}_{4}-\bkt{h}_{1}^{\rm Gibbs} \).
Let us construct a simple lower bound here.

We start from the representation (\ref{eq:S31rep}) of the entropy 
difference, and rewrite it as
\begin{eqnarray}
	S_{h'}[\rho_{3}^{\rm Gibbs}]-S_{h}[\rho_{1}^{\rm Gibbs}]
	&=&
	-k\sumin p_{i}\log\frac{p'_{i}}{p_{i}}
	\ret
	&=&
	k\sumin p_{i}\cbk{
	\rbk{1-\frac{p'_{i}}{p_{i}}}
	+\frac{p'_{i}}{p_{i}}-1-
	\log\frac{p'_{i}}{p_{i}}
	}
	\ret
	&=&
	k\sumin p_{i}\,g\!\!\rbk{\frac{p'_{i}}{p_{i}}},
	\label{eq:SSpg}
\end{eqnarray}
where
\begin{equation}
	g(x)=x-1-\log x.
	\label{eq:gx}
\end{equation}
Note that \( g(x)\ge0 \) for any \( x>0 \) and
\( g(x)>0 \) if \( x\ne1 \).

Let us first treat the simplest nontrivial case with \( n=3 \),
i.e., a three level system.
Let
\begin{equation}
	d_{1}=\ep_{3}-\ep_{1},\quad
	d_{2}=\ep_{3}-\ep_{2},\quad
	d'_{1}=\ep'_{3}-\ep'_{1},\quad
	d'_{2}=\ep'_{3}-\ep'_{2}.
	\label{eq:ddef}
\end{equation}
Then for \( p_{i} \) and  \( p'_{i} \) 
as in (\ref{eq:pi}) and (\ref{eq:pi'}), we have
\begin{equation}
	\frac{p_{3}}{p'_{3}}\frac{p'_{1}}{p_{1}}
	=\exp[-\beta d_{1}+\beta'd'_{1}],\quad
	\frac{p_{3}}{p'_{3}}\frac{p'_{2}}{p_{2}}
	=\exp[-\beta d_{2}+\beta'd'_{2}].
	\label{eq:pppp}
\end{equation}

Introducing vectors \( {\bf d}=(d_{1},d_{2}) \)
and
\( {\bf d}'=(d'_{1},d'_{2}) \),
we observe that
\begin{equation}
	\min_{\beta'}\abs{
	\beta{\bf d}-\beta'{\bf d}'
	}
	=
	\beta D({\bf d},{\bf d}'),
	\label{eq:min}
\end{equation}
where
\begin{equation}
	D({\bf d},{\bf d}')=
	|{\bf d}|
	\sqrt{
	1-
	\frac{({\bf d}\cdot{\bf d}')^2}
	{|{\bf d}|^2|{\bf d}'|^2}
	}.
	\label{eq:Ddd}
\end{equation}
Note that \( D({\bf d},{\bf d}') \) is vanishing if
\( \bf d \) and \( {\bf d}' \) are proportional,
and is strictly positive otherwise.
From (\ref{eq:min}) we see that
\begin{equation}
	|\beta d_{i}-\beta'd'_{i}|\ge
	\frac{\beta D({\bf d},{\bf d}')}{\sqrt{2}}
	\label{eq:dd}
\end{equation}
for at least 
one of \( i=1 \) or 2.
From this bound we can show that, 
at least for one of \( i=1,2,3 \),
we have
\begin{equation}
	\abs{\log\frac{p'_{i}}{p_{i}}}
	\ge
	\frac{\beta}{2\sqrt{2}}D({\bf d},{\bf d}').
	\label{eq:logpp}
\end{equation}
To see this, suppose the contrary, i.e.,
\( |\log({p'_{i}}/{p_{i}})|<
{\beta}D({\bf d},{\bf d}')/({2\sqrt{2}}) \) for all \( i \).
Then for \( i \) in (\ref{eq:dd})
\begin{eqnarray}
	\abs{\log\frac{p'_{i}}{p_{i}}}
	&=&
	\abs{\log\frac{p_{3}}{p'_{3}}\frac{p'_{i}}{p_{i}}
	+
	\log\frac{p'_{3}}{p_{3}}}
	\ret
	&=&
	\abs{-\beta d_{i}+\beta'd'_{i}
	+
	\log\frac{p'_{3}}{p_{3}}}
	\ret
	&\ge&
	|\beta d_{i}-\beta'd'_{i}|
	-
	\abs{\log\frac{p'_{3}}{p_{3}}}
	\ret
	&\ge&
	\frac{\beta}{2\sqrt{2}}D({\bf d},{\bf d}'),
	\label{eq:logproof}
\end{eqnarray}
which is a contradiction.
We have here used 
(\ref{eq:pppp}), (\ref{eq:dd}), the triangular inequality,
as well as the assumption
(\ref{eq:logpp}).

Let \( i \) be such that (\ref{eq:logpp}) is valid.
Since \( g(x) \) is increasing in \( x \) for
\( x>1 \) and decreasing in \( x \) for \( x<1 \),
we find that
\( g(p'_{i}/p_{i}) \) is bounded from below
by the smaller of
\( g(\exp[{\beta}D({\bf d},{\bf d}')/({2\sqrt{2}})]) \)
and
\( g(\exp[-{\beta}D({\bf d},{\bf d}')/({2\sqrt{2}})]) \).
Since the latter is smaller,
we have
\begin{equation}
	g\!\!\rbk{\frac{p'_{i}}{p_{i}}}
	\ge
	\frac{\beta}{2\sqrt{2}}D({\bf d},{\bf d}')
	+\exp\sqbk{-\frac{\beta}{2\sqrt{2}}D({\bf d},{\bf d}')}
	-1>0.
	\label{eq:gpp}
\end{equation}
By using the representation (\ref{eq:SSpg}),
we can bound the total entropy difference as
\begin{equation}
	S_{h'}[\rho_{3}^{\rm Gibbs}]-S_{h}[\rho_{1}^{\rm Gibbs}]
	\ge
	k\,p_{i}\,g\!\!\rbk{\frac{p'_{i}}{p_{i}}}
	\ge
	k\,p_{3}\,g\!\!\rbk{\frac{p'_{i}}{p_{i}}}.
	\label{eq:S31e}
\end{equation}
By combining this with (\ref{eq:S53}) and\footnote{
Although we can construct a lower bound for
\( S_{h}[\rho_{5}^{\rm Gibbs}]-S_{h'}[\rho_{3}^{\rm Gibbs}] \)
in the same manner,
it will depend on \( \beta' \) and won't be an explicit 
bound as it is.
Therefore we simply use the weak inequality
 (\ref{eq:S53}) for the second operation.
}
using (\ref{eq:gpp}), we get an explicit bound
for the entropy increase\footnote{If one prefers a bound which include only simple functions,
one may use
\begin{displaymath}
	S_{h}[\rho_{5}^{\rm Gibbs}]-S_{h}[\rho_{1}^{\rm Gibbs}]
	\ge
	\frac{ke^{-\beta\ep_{3}}}{z(\beta)}
	\times
	\cases{
	\frac{\beta^2}{24}\{D({\bf d},{\bf d}')\}^2
	&if \( \beta D({\bf d},{\bf d}')/(2\sqrt{2})\le(4/3)  \)\cr
	\frac{\beta}{2\sqrt{2}}D({\bf d},{\bf d}')-1
	&if \( \beta D({\bf d},{\bf d}')/(2\sqrt{2})\ge(4/3)  \),\cr
	}
\end{displaymath}
which is obtained from (\ref{eq:S51e}) through elementary estimate.
}
\begin{equation}
	S_{h}[\rho_{5}^{\rm Gibbs}]-S_{h}[\rho_{1}^{\rm Gibbs}]
	\ge
	\frac{ke^{-\beta\ep_{3}}}{z(\beta)}
	\cbk{
	\frac{\beta}{2\sqrt{2}}D({\bf d},{\bf d}')
	+\exp\sqbk{-\frac{\beta}{2\sqrt{2}}D({\bf d},{\bf d}')}
	-1
	}>0.
	\label{eq:S51e}
\end{equation}

We can use the same estimate to treat models with general \( n \).
Let \( M \) be the smallest integer which does not exceed \( n/3 \).
For each \( m=1,\ldots,M \), we choose 
three integers \( 1\le\ell_{1}(m)<\ell_{2}(m)<\ell_{3}(m)\le n \),
in such a way that each \( i=1,\ldots,n \) is chosen at most 
once\footnote{
A simple choice is \( \ell_{i}(m)=3(m-1)+i \).
}.
Then by using the previous estimate for the levels
\( \ep_{i} \), \( \ep'_{i} \) with 
\( i=\ell_{1}(m),\ell_{2}(m),\ell_{3}(m) \),
we get
\begin{equation}
	S_{h}[\rho_{5}^{\rm Gibbs}]-S_{h}[\rho_{1}^{\rm Gibbs}]
	\ge
	\sum_{m=1}^M
	\frac{ke^{-\beta\ep_{\ell_{3}(m)}}}{z(\beta)}
	\cbk{
	\frac{\beta}{2\sqrt{2}}D({\bf d}_{m},{\bf d}'_{m})
	+\exp\sqbk{-\frac{\beta}{2\sqrt{2}}D({\bf d}_{m},{\bf d}'_{m})}
	-1
	}>0,
	\label{eq:S51efull}
\end{equation}
where
\begin{equation}
	{\bf d}_{m}=(\ep_{\ell_{3}(m)}-\ep_{\ell_{1}(m)},
	\ep_{\ell_{3}(m)}-\ep_{\ell_{2}(m)}),
	\label{eq:dm}
\end{equation}
and
\begin{equation}
	{\bf d}'_{m}=(\ep'_{\ell_{3}(m)}-\ep'_{\ell_{1}(m)},
	\ep'_{\ell_{3}(m)}-\ep'_{\ell_{2}(m)}).
	\label{eq:d'm}
\end{equation}
It is clear that whenever the energy levels \( \ep_{i} \),
\( \ep'_{i} \) do not satisfy the condition (\ref{eq:hh'cond}),
one can choose \( \ell_{i}(m) \) in such a way that 
\( D({\bf d}_{m},{\bf d}'_{m})>0 \) for at least one \( m \).

Finally, we note that the entropy and the energy in 
the Gibbs state
for the Hamiltonian \( h \) at a general
inverse temperature \( \tbeta \) satisfies
the wellknown relation
\begin{equation}
	\frac{\partial}{\partial\tbeta}
	S_{h}(\rho_{\tbeta,h}^{\rm Gibbs})
	=
	k\tbeta\frac{\partial}{\partial\tbeta}
	\bkt{h}_{\tbeta,h}^{\rm Gibbs}.
	\label{eq:dSdh}
\end{equation}
Dividing this relation by \( -k\tbeta \),
and integrating it from \( \beta'' \) to \( \beta \),
one finds that
\begin{eqnarray}
	\bkt{h}_{5}^{\rm Gibbs}-\bkt{h}_{1}^{\rm Gibbs}
	&=&
	\int_{\beta''}^{\beta}d\tbeta
	\frac{1}{k\tbeta}
	\cbk{-
	\frac{\partial}{\partial\tbeta}
	S_{h}(\rho_{\tbeta,h}^{\rm Gibbs})}
	\ret
	&\ge&
	\frac{1}{k\beta}
	\int_{\beta''}^{\beta}d\tbeta
	\cbk{-
	\frac{\partial}{\partial\tbeta}
	S_{h}(\rho_{\tbeta,h}^{\rm Gibbs})}
	\ret
	&=&
	\frac{1}{k\beta}\{
	S_{h}[\rho_{5}^{\rm Gibbs}]-S_{h}[\rho_{1}^{\rm Gibbs}]
	\}.
	\label{eq:eandS}
\end{eqnarray}
By substituting (\ref{eq:S51efull})
and recalling (\ref{eq:h5h4}), we get our final estimate
for the irreversible work
\begin{equation}
	\bkt{h}_{4}-\bkt{h}_{1}^{\rm Gibbs}
	\ge
	\sum_{m=1}^M
	\frac{e^{-\beta\ep_{\ell_{3}(m)}}}{\beta\,z(\beta)}
	\cbk{
	\frac{\beta}{2\sqrt{2}}D({\bf d}_{m},{\bf d}'_{m})
	+\exp\sqbk{-\frac{\beta}{2\sqrt{2}}D({\bf d}_{m},{\bf d}'_{m})}
	-1
	}.
	\label{eq:final}
\end{equation}

We stress that the bound (\ref{eq:final}) is far from optimal.
In the crucial estimate (\ref{eq:min}), for example, 
we are simply looking for \( \beta' \) which minimizes the
distance between \( \beta{\bf d} \) and  \( \beta'{\bf d}' \).
But the true \( \beta' \) is determined from the 
condition (\ref{eq:h2=h3}) about the balance of energy,
which condition is not taken into account in our bound.

\bigskip

It is a pleasure to thank Ken Sekimoto for useful discussions
and for letting me know of the results in \cite{SSHT} before making 
them public.


\end{document}